\RequirePackage{fix-cm}
\documentclass{svjour3}                     
\smartqed  
\usepackage{xcolor,graphicx}
\usepackage[pdftex]{hyperref}
\usepackage{physics}
\usepackage{amsmath}
\usepackage{amssymb}
\usepackage{booktabs}
\usepackage[retainorgcmds]{IEEEtrantools}
\usepackage[ruled]{algorithm2e}
\newcommand{\beq}{\begin{equation}}
\newcommand{\eeq}{\end{equation}}
\newcommand{\ii}{\mathrm{i}}
\newcommand{\ee}{\mathrm{e}}

\newcommand{\nfft}{N_\mathrm{fft}}
\newcommand{\nblk}{N_\mathrm{blk}}

\newcommand{\nwin}{N_\mathrm{win}}
\newcommand{\nwinf}{N_{\mathrm{win},f_k}}

\begin{document}

\title{Adaptive spectral proper orthogonal decomposition of {broadband-}tonal flows
}


\author{Brandon C. Y. Yeung         \and
        Oliver T. Schmidt 
}


\institute{B. C. Y. Yeung \at
              Department of Mechanical and Aerospace Engineering, Jacobs School of Engineering, University of California San Diego, 9500 Gilman Drive, La Jolla, CA 92093-0411, USA \\
              \email{byeung@ucsd.edu}           
           \and
           O. T. Schmidt \at
              Department of Mechanical and Aerospace Engineering, Jacobs School of Engineering, University of California San Diego, 9500 Gilman Drive, La Jolla, CA 92093-0411, USA \\
              \email{oschmidt@ucsd.edu}
}

\date{Received: date / Accepted: date}

\maketitle

\begin{abstract}
An adaptive algorithm for spectral proper orthogonal decomposition (SPOD) of mixed broadband-tonal turbulent flows is developed. Sharp peak resolution at tonal frequencies is achieved by locally minimizing bias {of the spectrum}. Smooth spectrum estimates of broadband regions are achieved by locally reducing variance {of the spectrum}. The method utilizes multitaper estimation with sine tapers. An iterative criterion based on modal convergence is introduced to enable the SPOD to adapt to spectral features. For tonal flows, the adaptivity is controlled by a single user input; for broadband flows, a constant number of sine tapers is recommended without adaptivity. The discrete version of Parseval's theorem for SPOD is stated. Proper normalization of the tapers ensures that Parseval's theorem is satisfied in expectation. Drastic savings in computational complexity and memory usage are facilitated by two aspects: (i) sine tapers, which permit \textit{post hoc} windowing of a single Fourier transform; and (ii) time-domain lossless compression using a QR or eigenvalue decomposition. Sine-taper SPOD is demonstrated on time-resolved particle image velocimetry (TR-PIV) data from an open cavity flow \cite{ZhangEtAl2020ExpFluids} and high-fidelity large-eddy simulation (LES) data from a round jet \cite{BresEtAl2018JFM}, with and without adaptivity. For the tonal cavity flow, the adaptive algorithm outperforms Slepian-based multitaper SPOD in terms of {variance and local bias of the spectrum}, mode convergence, and memory usage. {The tonal frequencies associated with the Rossiter instability are accurately identified.} For both the tonal cavity and the broadband jet flows, results comparable to or better than those from standard SPOD based on Welch's overlapped segment averaging are obtained with up to 75\% fewer snapshots, {including similar convergence of the Rossiter modes and Kelvin-Helmholtz wavepacket structures for the cavity and jet examples, respectively}. {Drawing from these examples, we establish best practices.}

\end{abstract}

\section{Introduction}\label{sec:intro}
Owing to its stochastic nature, turbulence has been studied using a variety of data-driven statistical approaches. Among them are dynamic mode decomposition (DMD, \cite{Schmid2010JFM}) and proper orthogonal decomposition (POD, \cite{Lumley1970AP}), the latter of which forms the subject of this paper. POD in the most general form seeks a set of orthogonal modes that optimally represent the second-order space-time statistics of turbulent flows. For wide-sense stationary flows, space-time POD was later specialized by adding a temporal Fourier transform of the data, before a space-only POD is performed per frequency (see \cite{GlauserEtAl1987TSF} for an early example). This procedure is termed spectral POD (SPOD, \cite{PicardDelville2000HFF}). Since SPOD modes are monochromatic and coherent in both space and time, they can be interpreted as coherent structures. Convergence of SPOD modes and their modal energies often demands a large number of realizations of the Fourier transform. Collecting sufficient samples can be experimentally, numerically, and literally costly.

In the standard implementation of SPOD \cite{TowneEtAl2018JFM,SchmidtEtAl2018JFM,SchmidtColonius2020AIAAJ}, the samples are obtained using Welch estimation \cite{Welch1967IEEE}. The time-domain data are first segmented into overlapping blocks, each of which is then Fourier transformed. The block size is inversely related to the frequency resolution of the spectrum. For the sample size to be large, each block must be short, which lowers resolution and increases bias. Conversely, long blocks raise resolution, but make fewer samples available, increasing the variance of the estimate. More precise control of this trade-off can be achieved with Thomson's multitaper estimator \cite{Thomson1982IEEE}, recently adopted for SPOD by \cite{Schmidt2022TCFD}. Rather than segmenting the data, Thomson's multitaper windows them using a family of mutually orthogonal tapers, called discrete prolate spheroidal sequences (DPSS, also known as Slepians, \cite{SlepianPollak1961BSTJ,Slepian1978BSTJ}), that render the Fourier realizations approximately uncorrelated with each other. As each Fourier transform spans the entire data, spectral resolution is maximized. Independent of resolution, the variance and bias of the estimate are controlled via the number and bandwidth of the tapers. The DPSS are the family of tapers that maximize spectral concentration. Thomson also proposed adaptive weighting of the windowed Fourier transforms to offer additional protection against spectral leakage \cite{Thomson1982IEEE}. However, Thomson's multitaper estimator offers scant protection from local bias \cite{RiedelSidorenko1995IEEE}, a phenomenon that causes discrete tones in the spectrum to become smeared or flattened, though improvements were later made by \cite{PrietoEtAl2007GJI}. In the context of SPOD, \cite{Schmidt2022TCFD} found that Thomson's multitaper also incurs dramatically higher computational cost and memory usage relative to the Welch estimator and, as a compromise, suggested a hybrid multitaper-Welch approach.

Our overarching goal is to formulate an SPOD algorithm that combines strong variance reduction with low bias, and that is computationally efficient. Such an algorithm would be well-suited to estimating spectra that include tones due to hydrodynamic instabilities, resonances, and exogenous inputs. It would also be able to handle large data, such as those produced by high-fidelity simulations. To this end, we propose an adaptive multitaper SPOD that uses sine functions as tapers, with the aim of achieving better local bias protection and lower memory usage than DPSS-taper SPOD, while ensuring reasonable convergence of modes at all frequencies. Introduced by \cite{RiedelSidorenko1995IEEE} for one-dimensional signal processing, sine tapers optimally minimize local bias. The properties of sine functions also allow an arbitrary number of sine-tapered Fourier realizations to be assembled from a single Fourier transform. The reduction in computational effort this affords may be immaterial to one-dimensional signal processing, but is crucial in the context of SPOD. To further mitigate computational complexity and peak memory consumption, we demonstrate lossless compression of the time-domain data. When a spectrum contains tones, frequency-dependent estimation parameters enable the bias and variance to adapt to the spectrum. {Riedel and Sidorenko} \cite{RiedelSidorenko1995IEEE} derived an expression for the optimum number of sine tapers, in an expected error sense, to apply as a function of frequency, with the caveat that the expression depends on the unobservable true spectrum. {Barbour and Parker} \cite{BarbourParker2014CAGEO} substituted the true spectrum with an estimated spectrum, then determined the taper number by iteratively updating the estimate. {Hansson} \cite{Hansson1999IEEE} obtained frequency-dependent weights for tapered Fourier transforms, including those computed using sine tapers, through an iterative procedure that also requires knowledge of the true spectrum. The application of sine tapers to spectral modal decomposition offers a novel pathway to realizing adaptivity. Unlike signal processing, SPOD accounts for spatial coherence. It requires mode shapes to be well-converged so that they may be employed for physical discovery and reduced-order modeling (ROM). Convergence of the mode shapes is in turn coupled to that of the spectrum. We propose an adaptive algorithm that protects tones in the spectrum from bias by leveraging the spatial information inherent in SPOD. Using mode convergence, the algorithm determines a taper number for each frequency such that a smaller number is applied to tones than to broadband turbulence. We demonstrate adaptive sine-taper SPOD on time-resolved particle image velocimetry (TR-PIV) data of an open cavity flow \cite{ZhangEtAl2020ExpFluids}, which contain noise and mix tones with broadband turbulence, and show that the judicious selection of a single parameter, the similarity tolerance, suffices to generate a spectrum with low bin-to-bin variance and sharp tones, and smooth modes at a broad range of frequencies. Furthermore, we apply non-adaptive sine-taper SPOD to high-fidelity large-eddy simulation (LES) data of a turbulent round jet \cite{BresEtAl2018JFM} to discuss the advantages of the technique for non-tonal, broadband flows.

\begin{figure}
    \centering
    \includegraphics{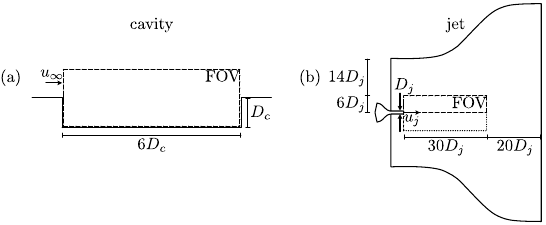}
    \caption{{2D representations of the geometries of the cavity PIV and jet LES data. The field of view (FOV) of each case is indicated by dashed lines. The cavity has a depth of $D_c$ and a freestream velocity of $u_\infty$. The jet has a nozzle diameter of $D_j$ and a nozzle exit velocity of $u_j$. For the jet case, since only the axisymmetric component is considered, the region marked by dotted lines is redundant}}
    \label{fig:schematic}
\end{figure}

\begin{figure}[h]
    \centering
    \includegraphics{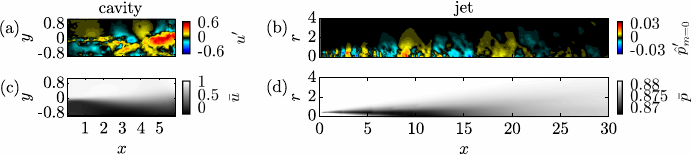}
    \caption{Instantaneous fluctuations (a,b) and mean (c,d) of: (a,c) streamwise velocity component of the cavity PIV data; (b,d) axisymmetric pressure component of the jet LES data. For the remainder of the paper, for clarity of visualizations, we will focus on subdomains but have confirmed that all conclusions hold for the full domain}
    \label{fig:dataoverview}
\end{figure}

The remainder of the paper is organized as follows. The standard SPOD algorithm is briefly summarized in Sect.~\ref{sec:Spod}. Sect.~\ref{sec:multitaper} introduces multitaper SPOD with sine tapers. Acceleration of SPOD via lossless compression is developed in Sect.~\ref{sec:compression}. Sect.~\ref{sec:adaptive} introduces the adaptive algorithm. Applications of sine-taper SPOD to the cavity and jet flows, with and without adaptivity, are presented in Sects.~\ref{sec:cavity} and \ref{sec:jet} respectively and discussed in Sect.~\ref{sec:discussion}. The parameters of the cavity and jet data are summarized in Table~\ref{tab:dataoverview}. {Their geometries are illustrated in Fig.~\ref{fig:schematic},} and their flow fields are visualized in Fig.~\ref{fig:dataoverview}. Appendix~\ref{sec:appScaling} reports computational performance data. A strategy to tackle excessive time-to-solution is outlined in Appendix~\ref{sec:appLossy}.
\begin{table}[h]
    \centering
    \caption{Parameters of the cavity PIV and jet LES data. {For the cavity, the time step is expressed in seconds (see Sect.~\ref{sec:cavity}). For the jet, the time step is non-dimensionalized by the nozzle diameter, $D_j$, and nozzle exit velocity, $u_j$ (see Sect.~\ref{sec:jet}).} Only the axisymmetric component of the jet data is used}
    \label{tab:dataoverview}
    \begin{tabular}{@{}lcccccc@{}}\toprule
        Case & $M$ & $\vb{q}$ & $\Delta t$ & $N_t$ & $N_x$ & $N_y$ or $N_r$ \\\midrule
        Cavity PIV & 0.6 & $[\vb{u},\vb{v}]^\mathrm{T}$ & {$6.25\times10^{-5}$ s} & 16000 & 156 & 55 \\
        Jet LES & 0.9 & $[\hat{\vb*{\rho}}, \hat{\vb{u}}_x, \hat{\vb{u}}_r, \hat{\vb{u}}_\theta, \hat{\vb{T}}]^\mathrm{T}$ & {$0.18 D_j/u_j$} & 10000 & 656 & 138 \\\bottomrule
    \end{tabular}
\end{table}

\section{Adaptive multitaper SPOD}\label{sec:method}
\subsection{SPOD}\label{sec:Spod}
We begin with an outline of the SPOD algorithm. Given $N_t$ snapshots $\vb{q}_{t_j}\in\mathbb{C}^{{N_\mathrm{dof}}}$, $j=1,\ldots,N_t$, of a statistically stationary flow sampled at {$N_\mathrm{dof}$} points, 
where {$N_\mathrm{dof}$ is the number of degrees of freedom (the product of the number of spatial locations and the number of variables)} recorded, the Reynolds decomposition
\beq
\vb{q}_{t_j} = \bar{\vb{q}} + \vb{q}'_{t_j}
\eeq
partitions $\vb{q}$ into the long-time mean, $\bar{\vb{q}}$, and the fluctuations, $\vb{q}'$.  In general the snapshots may be complex, for example after Fourier transforms in one or more homogeneous directions. The fluctuations are assembled in the data matrix
\beq
\vb{Q}' = \mqty[\vb{q}'_{t_1}, & \ldots, & \vb{q}'_{t_{N_t}}] \in\mathbb{C}^{{N_\mathrm{dof}}\times N_t}.
\eeq
In the following, we will consider fluctuation quantities only and drop the $(\cdot)'$ notation.

At the core of SPOD is the eigendecomposition of the weighted cross-spectral density (CSD) matrix, $\vb{S}_{f_k} =  \hat{\vb{X}}_{f_k}\hat{\vb{X}}^*_{f_k}$, where $\hat{\vb{X}}_{f_k}=\sqrt{\Delta t/N}\vb{W}^\frac{1}{2}\hat{\vb{Q}}_{f_k}$, $\Delta t$ is the time step between snapshots, $\vb{W}$ is a diagonal weight matrix, and the columns of $\hat{\vb{Q}}_{f_k}$ comprise $N$ realizations of the temporal Fourier transform of $\vb{Q}$ at frequency $f_k$. At each frequency, the eigendecomposition
\beq
\vb{S}_{f_k}\vb{\Theta}_{f_k} = \vb{\Theta}_{f_k}\vb{\Lambda}_{f_k}
\eeq
provides the SPOD modal energies, $\mathrm{diag}(\vb{\Lambda}_{f_k})$, and the modes, $\vb{\Phi}_{f_k}=\vb{W}^{-\frac{1}{2}}\vb{\Theta}_{f_k}$. Inclusion of the weight ensures that the modes are optimal in the spatial inner product
\beq
\expval{\hat{\vb{q}}_1,\hat{\vb{q}}_2}_{\vb{x}} = \hat{\vb{q}}^*_2 \vb{W} \hat{\vb{q}}_1. \label{eq:innerProd}
\eeq

If ${N_\mathrm{dof}}>N$, as is typical in fluid flow applications, the SPOD is more efficiently computed using the method-of-snapshots \cite{Sirovich1987QAM1}. Rather than explicitly constructing then diagonalizing the {$N_\mathrm{dof}\times N_\mathrm{dof}$} CSD matrix, we solve the $N\times N$ eigendecomposition,
\beq
\hat{\vb{X}}^*_{f_k}\hat{\vb{X}}_{f_k}\vb{\Psi}_{f_k} = \vb{\Psi}_{f_k}\vb{\Lambda}_{f_k}, \label{eq:sirovichEvp}
\eeq
to obtain the same nonzero eigenvalues, $\mathrm{diag}(\vb{\Lambda}_{f_k})$, and modes, 
\beq
\vb{\Phi}_{f_k}=\vb{W}^{-\frac{1}{2}}\hat{\vb{X}}_{f_k}\vb{\Psi}_{f_k}\vb{\Lambda}^{-\frac{1}{2}}_{f_k}, \label{eq:sirovichEvpB}
\eeq
as before. For details on the SPOD algorithm, we refer the reader to \cite{TowneEtAl2018JFM,SchmidtTowne2019CPC,SchmidtColonius2020AIAAJ}.

\subsection{Multitaper estimation}\label{sec:multitaper}
Estimation of the CSD matrix requires assembling $N>1$ independent realizations of the Fourier mode (coefficient), $\hat{\vb{q}}^{(n)}_{f_k}$, $n=1,\ldots,N$, at each frequency. The classic Welch method \cite{Welch1967IEEE} produces these realizations by segmenting a single time series into $N=\nblk$ partially overlapping blocks of length $\nfft$, before taking the windowed fast Fourier transform (FFT) of each block. The resulting spectral estimates have a normalized half-bandwidth of $\delta f=1/\nfft$. For this estimator, the variance-bias trade-off necessitates a compromise between maximizing $N$, which reduces variance in the estimates, and maximizing $\nfft$, which improves frequency resolution. In contrast, the multitaper estimator \cite{Thomson1982IEEE} produces $N$ independent realizations using $N=\nwin$ mutually orthogonal windows, or tapers, $\vb{v}^{(n)}=\mqty[v^{(n)}_1, & \ldots, & v^{(n)}_{\nfft}]$, then performing and storing an FFT for each window:
\beq
\hat{\vb{q}}^{(n)}_{f_k} = \sum_{j=0}^{\nfft-1} v^{(n)}_{j+1}\vb{q}_{t_{j+1}}\ee^{-\ii2\pi jk/{\nfft}}, \ k=0,\ldots,\nfft-1,\ n=1,\ldots,\nwin. \label{eq:windowedFFT}
\eeq
The choice of tapers in the first application of multitaper estimation to SPOD \cite{Schmidt2022TCFD} were the DPSS. These uniquely maximize the spectral concentration of the estimates within a given bandwidth and, lacking an analytical form, must be computed from a Hermitian eigendecomposition. {Determining an optimal choice of $\nwin$ forms the subject of Sect.~\ref{sec:adaptive}.}

In controlling the variance-bias trade-off a different objective is to minimize the bias due to local variation of the true spectrum within the finite bandwidth of the estimator. By expanding the true spectrum in a Taylor series, it can be shown (see e.g. \cite{Thomson1982IEEE}) that this local bias is dominated by a term proportional to $\int_{-\frac{1}{2\Delta t}}^{\frac{1}{2\Delta t}} \abs{\hat{v}(f)}^2 f^2 \dd f$, where $\hat{v}$ is the discrete-time Fourier transform of the taper, $\vb{v}$. The tapers that uniquely minimize this term are referred to as the minimum bias tapers \cite{RiedelSidorenko1995IEEE}. Like the DPSS, they are the solutions to a Hermitian eigenvalue problem (EVP), but can be closely approximated by the set of scaled sinusoids \cite{RiedelSidorenko1995IEEE},
\beq
v_j^{(n)} = \sqrt{\frac{2}{\nfft+1}} \sin\frac{\pi n j}{\nfft+1},\quad j=1,\ldots,\nfft,\quad n=1,\ldots,\nwin, \label{eq:sine}
\eeq
where $j$ is the discrete time index, and $n$ the taper index. The $\sqrt{2/(\nfft+1)}$ factor normalizes each sine taper such that $\sum_j (v_j^{(n)})^2=1$. This ensures that the FFT in Eq.~\eqref{eq:windowedFFT} satisfies Parseval's theorem \cite{PercivalWalden1993Cambridge}, which also relates SPOD eigenvalues to average power. With this normalization, the discrete version of Parseval's theorem for SPOD guarantees that the eigenvalues at each frequency sum to the spatially-integrated power spectral density (PSD); the integral of the latter over frequency recovers the average power of the zero-mean stationary process. For discrete time and frequency, this relationship can be expressed as
\begin{IEEEeqnarray}{rCl}
& \Delta f\sum_{k=0}^{\nfft-1}\sum_{i=1}^{\nwin} E\qty{\lambda_{i,f_k}} = \norm{\vb*{\sigma}}^2_{\vb{x}} = \frac{1}{N_t} \sum_{j=1}^{N_t} E\qty{\norm{\vb{q}_{t_j}}^2_{\vb{x}}} & \nonumber\\
& \text{({Parseval's theorem for SPOD}),} & \label{eq:parseval}
\end{IEEEeqnarray}
where $\Delta f=1/(\nfft\Delta t)$ is the frequency bin width, $\lambda$ the estimated eigenvalues, $E\qty{\cdot}$ the expectation operator, $\vb*{\sigma}$ the pointwise true standard deviation, and $\norm{\vb{q}}^2_{\vb{x}}=\expval{\vb{q},\vb{q}}_{\vb{x}}$. Given a finite number of samples---as is always the case when dealing with data---the identity can instead be approximated as
\beq
\Delta f\sum_{k=0}^{\nfft-1}\sum_{i=1}^{\nwin} \lambda_{i,f_k} \approx \frac{1}{N_t} \sum_{j=1}^{N_t} \norm{\vb{q}_{t_j}}^2_{\vb{x}}, \label{eq:parsevalApprox}
\eeq
which is exact only for the rectangular window, with $v_j=1/\sqrt{\nfft}$. Conservation of power, i.e., properties~\eqref{eq:parseval} and ~\eqref{eq:parsevalApprox}, simplify comparisons between eigenvalues computed using different spectral estimation parameters.

Relative to DPSS tapers, which by construction are optimal in spectral concentration, sine tapers have slightly lower spectral concentration \cite{RiedelSidorenko1995IEEE}. With DPSS tapers, customarily the user supplies the normalized half-bandwidth, $\delta f$, from which the number of usable tapers is derived as $\nwin=\lfloor 2\delta f\nfft \rfloor-1$ \cite{Thomson1982IEEE}. With sine tapers, the user instead directly supplies $\nwin$, which translates to a half-bandwidth of $\delta f=(\nwin+1)/[2(\nfft+1)]$ \cite{WaldenEtAl1995Biomet}. In DPSS-taper SPOD \cite{Schmidt2022TCFD}, the Fourier realizations are uniformly weighted with weight $\mu_n=1/\nwin$, $n=1,\ldots,\nwin$. In sine-taper SPOD, we follow the recommendation of \cite{RiedelSidorenko1995IEEE} and apply nonuniform weighting to each Fourier realization. The weights are given by
\beq
\mu_n = \frac{6}{4\nwin^3+3\nwin^2-\nwin}[\nwin^2-(n-1)^2], \quad n=1,\ldots,\nwin, \label{eq:sinParabWgt}
\eeq
and satisfy $\sum_{n=1}^{\nwin}\mu_n=1$. They penalize higher-order tapers that possess larger bandwidths. With these weights, we now define the modified matrix of Fourier realizations as
\beq
\hat{\vb{X}}_{f_k} = \sqrt{\Delta t}\vb{W}^\frac{1}{2}\hat{\vb{Q}}_{f_k} \mathrm{diag}([\mu_1,\ldots,\mu_{\nwin}])^\frac{1}{2}. \label{eq:dataParabWgt}
\eeq

The simple analytical form of the sine tapers in Eq. \eqref{eq:sine} affords a significant computational simplification \cite{RiedelSidorenko1995IEEE} via the identity
\beq
\hat{\vb{q}}^{(n)}_{f_k} = \sqrt{\frac{2}{\nfft+1}} \frac{\hat{\vb{q}}_{f_{k-n\Delta k}}-\hat{\vb{q}}_{f_{k+n\Delta k}}}{2\ii},\quad n=1,\ldots,\nwin, \label{eq:ctrDiff}
\eeq
where $\hat{\vb{q}}_{f_{k-n\Delta k}}$ and $\hat{\vb{q}}_{f_{k+n\Delta k}}$ are the unwindowed Fourier modes at frequency indices $k-n\Delta k$ and $k+n\Delta k$, and $\Delta k=\nfft/[2(\nfft+1)]\approx 0.5$ for $\nfft\gg1$.\footnote{In our experience the quality of this approximation only becomes relevant for values of $\nfft$ that are too low to be suitable for multitaper SPOD; for all practical purposes, it is appropriate to work with $\Delta k=1/2$.} The unwindowed modes, $\hat{\vb{q}}_{f_{k\pm n\Delta k}}$, for both even and odd $n$ are obtained by zero-padding the snapshots in the time domain (see e.g. \cite{OppenheimSchafer2010Pearson}) to length $2\nfft$, then performing the FFT:
\beq
\hat{\vb{q}}_{f_k}=\sum_{j=0}^{2\nfft-1} \vb{q}_{t_{j+1}}\ee^{-\ii2\pi jk/(2{\nfft})}, \ k=0,\ldots,2\nfft-1, \label{eq:paddedFft}
\eeq
where $\vb{q}_{t_{j+1}} = \vb{0}$ for $j=\nfft,\ldots,2\nfft-1$. Since only one FFT needs to be computed and stored, regardless of the number of tapers, a significant computational advantage is attained. Thereafter the windowed Fourier realizations, $\hat{\vb{q}}^{(n)}_{f_k}$, can be recovered using Eq. \eqref{eq:ctrDiff} by sampling from $\hat{\vb{q}}_{f_k}$ in Eq. \eqref{eq:paddedFft}. In \cite{Schmidt2022TCFD}, memory considerations motivated a hybrid multitaper-Welch SPOD, with a user-specified $\nfft<N_t$ and $\nblk>1$. The central difference identity in Eq. \eqref{eq:ctrDiff} largely neutralizes such concerns, as the scaling experiments in Appendix~\ref{sec:appScaling} attest. For this work, we set $\nfft=N_t$, both maximizing frequency resolution and simplifying user inputs. We will revisit this decision in Sects.~\ref{sec:cavityConstNwin} and \ref{sec:discussion}.

\subsection{Lossless compression}\label{sec:compression}
For large data, even with the techniques of Sects.~\ref{sec:Spod} and \ref{sec:multitaper}, sine-taper SPOD can be prohibitively slow.
To accelerate the SPOD, we propose to losslessly compress the weighted data matrix, assumed to be full-rank, using its QR factorization
\beq
\vb{X} = \vb{W}^\frac{1}{2}\vb{Q} = \vb{U}\widetilde{\vb{X}}, \label{eq:QR}
\eeq
where $\vb{U}\in\mathbb{C}^{{N_\mathrm{dof}}\times N_t}$ is an orthonormal basis for the column space of $\vb{X}$, and $\widetilde{\vb{X}}\in\mathbb{C}^{N_t\times N_t}$ is the compressed data matrix. For data whose ${N_\mathrm{dof}}\gg N_t$, it can be more efficient to orthogonalize $\vb{X}$ using the method-of-snapshots via the EVP
\beq
\vb{X}^*\vb{X}\vb{V} = \vb{V}\vb{L}, \label{eq:methOfSnapA}
\eeq
where $\vb{V}\in\mathbb{C}^{N_t\times N_t}$ contains the eigenvectors, and $\vb{L}\in\mathbb{R}^{N_t\times N_t}$ contains the eigenvalues on its diagonal. The orthonormal basis and compressed data matrix are recovered as
\beq
\vb{U} = \vb{X}\vb{V}\vb{L}^{-\frac{1}{2}} \quad\text{and}\quad \widetilde{\vb{X}}=\vb{L}^\frac{1}{2}\vb{V}^*, \label{eq:methOfSnapB}
\eeq
respectively. In practice, it is advantageous to replace the QR in Eq.~\eqref{eq:QR} with the method-of-snapshots in Eqs.~\eqref{eq:methOfSnapA} and \eqref{eq:methOfSnapB} above a threshold of ${N_\mathrm{dof}}\gtrsim 8N_t$. As an added benefit, compression of the snapshot matrix further reduces the memory consumption of the FFT in Eq.~\eqref{eq:paddedFft}. It can also be deduced from the dimensions of $\widetilde{\vb{X}}$ in Eqs.~\eqref{eq:QR} and \eqref{eq:methOfSnapB} that compression renders the overall wall time of the SPOD largely independent of {$N_\mathrm{dof}$}; this is exemplified in Appendix~\ref{sec:appScaling}.

The remainder of sine-taper SPOD proceeds almost unchanged from the steps in Sects.~\ref{sec:Spod} and \ref{sec:multitaper}.
We assemble the compressed Fourier realizations, $\hat{\widetilde{\vb{X}}}_{f_k}$, at frequency $f_k$ according to Eqs.~\eqref{eq:paddedFft}, \eqref{eq:ctrDiff}, and \eqref{eq:dataParabWgt}, then solve the $\nwin\times\nwin$ EVP
\beq
\hat{\widetilde{\vb{X}}}^*_{f_k} \hat{\widetilde{\vb{X}}}_{f_k} \vb{\Psi}_{f_k} = \vb{\Psi}_{f_k} \vb{\Lambda}_{f_k}. \label{eq:sirovichEvpComp}
\eeq
Since $\vb{U}$ is a complete orthonormal basis,
\beq
\hat{\vb{X}}^*_{f_k} \hat{\vb{X}}_{f_k} = \hat{\widetilde{\vb{X}}}^*_{f_k} \hat{\widetilde{\vb{X}}}_{f_k},
\eeq
i.e., the compression is lossless. The uncompressed EVP in Eq.~\eqref{eq:sirovichEvp} and its compressed counterpart in Eq.~\eqref{eq:sirovichEvpComp} thus yield identical SPOD modal energy and expansion coefficients, $\vb{\Lambda}_{f_k}$ and $\vb{\Psi}_{f_k}$, respectively. The SPOD modes are recovered from the compressed data as
\beq
 \vb{\Phi}_{f_k} = \vb{W}^{-\frac{1}{2}} \vb{U} \widetilde{\vb{\Theta}}_{f_k}, \quad\text{with}\quad \widetilde{\vb{\Theta}}_{f_k} = \hat{\widetilde{\vb{X}}}_{f_k} \vb{\Psi}_{f_k} \vb{\Lambda}^{-\frac{1}{2}}_{f_k}. \label{eq:decompressSpod}
\eeq

If the computational effort remains excessive, the SVD version of the lossless compression above may readily be used for lossy compression through a rank-$d$ truncation, with $d<N_t$. The optimality of the truncation is guaranteed by the properties of SVD. We accomplish this by retaining the first $d$ rows of $\widetilde{\vb{X}}$ in Eq.~\eqref{eq:methOfSnapB}. Appendix~\ref{sec:appLossy} offers an example in which compression is achieved by retaining the first $d=1000$ rows, and that shows an 87\% reduction in computation time, while the spectrum and modes remain visually indistinguishable from the losslessly compressed results.

Though we have proposed the application of spatial compression in the context of sine-taper SPOD, compression is agnostic to the choice of spectral estimator, and could be applied to DPSS-taper SPOD, as well as to standard Welch SPOD.


\subsection{Adaptivity}\label{sec:adaptive}
Two distinctive features of sine tapers enable the bandwidth to vary with frequency \cite{RiedelSidorenko1995IEEE}. They are: (i) the bandwidth depends solely on $N_\mathrm{win}$; and (ii) an arbitrary number of Fourier realizations can be sampled from one precomputed FFT. For any frequency $f_k$, the bandwidth can be locally broadened or narrowed by raising or lowering the number of tapers, $\nwinf$. In principle, $\nwinf$ can vary arbitrarily to locally achieve the desired balance between variance and bias. In practice, however, bin-to-bin discontinuities in $\nwinf$ may lead to undesired artifacts. To mitigate such artifacts, a best practice proposed by \cite{BarbourParker2014CAGEO} limits the bin-to-bin difference in $\nwinf$ to one, that is,
\beq
\abs{N_{\mathrm{win},f_k} - N_{\mathrm{win},f_{k\pm 1}} } \le 1. \label{eq:nwinConstrain}
\eeq
Despite this constraint, for temporally finely-resolved data and $\nfft=N_t$, the resulting high frequency resolution of the spectrum permits $\nwinf$ to vary rapidly over a small frequency interval---a desirable trait.

To determine $\nwinf$, we exploit the fact that SPOD modes must be well-converged in order to be utilized successfully for physical discovery and ROM. We therefore propose to use the convergence of the leading SPOD mode, $\vb*{\phi}_{1,f_k}$, between iterations to gauge the $\nwinf$ required at each frequency. At the $i$th iteration, we compute the sine-taper SPOD with $\nwin^{(i)}$ tapers, then evaluate the similarity \cite{SemeraroEtAl2016AIAA} between the leading modes at the $i$th and $(i-1)$th iterations:
\beq
\alpha^{(i)}_{f_k} = \abs{\qty(\vb*{\phi}^{(i)}_{1,f_k})^* \vb{W} \vb*{\phi}^{(i-1)}_{1,f_k}}. \label{eq:similarity}
\eeq
The weight, $\vb{W}$, is the same as that in the inner product of Eq.~\ref{eq:innerProd}, thus ensuring the similarity measure is consistent with the energy norm of the SPOD. Since SPOD modes are normalized to unit energy, by the Cauchy-Schwarz inequality, $\alpha^{(i)}_{f_k}\le1$. This implies that $\alpha^{(i)}_{f_k} \to 1$ as $\vb*{\phi}^{(i)}_{1,f_k} \to \vb*{\phi}^{(i-1)}_{1,f_k}$. Using Eq.~\eqref{eq:decompressSpod}, the similarity can be computed entirely in the compressed subspace as
\beq
\alpha^{(i)}_{f_k} = \abs{\qty(\widetilde{\vb*{\theta}}^{(i)}_{1,f_k})^* \widetilde{\vb*{\theta}}^{(i-1)}_{1,f_k}}. \label{eq:similarityCompress}
\eeq
We consider $\vb*{\phi}^{(i)}_{1,f_k}$, or equivalently $\widetilde{\vb*{\theta}}^{(i)}_{1,f_k}$, converged if $\alpha^{(i)}_{f_k}$ satisfies
\beq
1-\alpha^{(i)}_{f_k} \le tol, \label{eq:criterion}
\eeq
where $tol$ is the similarity tolerance.

One of our goals is to protect tones from local bias, which requires $\nwinf$ to be minimized. We achieve this by initializing the iterations with a small $\nwin$, then incrementally increasing it. In each subsequent iteration, we fix the taper number for all frequencies that either meet the convergence criterion~\eqref{eq:criterion}, or cannot accept more tapers due to constraint~\eqref{eq:nwinConstrain}. The frequencies that converge most rapidly are thus windowed using small numbers of tapers, with narrow bandwidths. Such frequencies are associated with tones or dominant hydrodynamic instabilities characterized by low-rankness \cite{SchmidtEtAl2018JFM}, i.e., a large separation between the leading and suboptimal SPOD eigenvalues. Where the spectrum is broadband, the prevalent flow structures are expected to vary smoothly as a function of frequency. We leverage this for statistical convergence by widening the bandwidth in regions of broadband turbulence.

\begin{figure}
    \centering
    \includegraphics{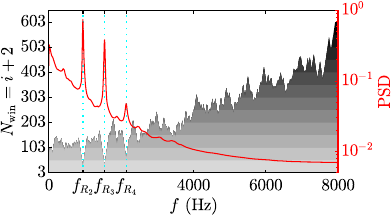}
    \caption{The taper numbers, $\nwin$, of the cavity flow generated iteratively by the adaptive SPOD algorithm, shown for every 50 iterations starting from $i=1$. Darker contours denote higher iteration counts. The volume-integrated PSD, in red, is superimposed. The second, third, and fourth Rossiter frequencies are marked by dotted cyan lines}
    \label{fig:cavity_nwin_iter}
\end{figure}

\begin{algorithm}    
    \SetAlgoLined
    \DontPrintSemicolon
    \SetKwComment{comment}{\% }{}
    \KwIn{$\vb{Q}$, $\vb{W}$, and $tol$.}
    \KwOut{$\vb{\Lambda}_{f_k}$, $\vb{\Phi}_{f_k}$, and $\nwinf$ for $k=0,\ldots,N_t-1$.}
    Obtain the unwindowed Fourier modes, $\hat{\vb{q}}_{f_k}$ \eqref{eq:paddedFft}.\;
    Initialize $i \gets 0$ and $\nwin^{(i)} \gets 2$.\;
    \While{not all $\nwinf$ fixed}{
        $i \gets i+1$.\;
        $\nwin^{(i)} \gets \nwin^{(i-1)}+1$.\;
        \For{all $k$ for which $\nwinf$ not fixed}{
            Assemble $\hat{\vb{X}}^{(i)}_{f_k}$ for $\nwin^{(i)}$ tapers from $\hat{\vb{q}}_{f_k}$ \eqref{eq:ctrDiff}, and solve ($\hat{\vb{X}}^{(i)}_{f_k})^*\hat{\vb{X}}^{(i)}_{f_k}\vb{\Psi}^{(i)}_{f_k} = \vb{\Psi}^{(i)}_{f_k}\vb{\Lambda}^{(i)}_{f_k}$ for $\vb{\Lambda}^{(i)}_{f_k}$ and $\vb{\Psi}^{(i)}_{f_k}$ \eqref{eq:sirovichEvp}. \comment*{EVP}
            $\vb{\Phi}^{(i)}_{f_k} \gets \vb{W}^{-\frac{1}{2}}\hat{\vb{X}}^{(i)}_{f_k}\vb{\Psi}^{(i)}_{f_k}(\vb{\Lambda}^{(i)}_{f_k})^{-\frac{1}{2}}$ \eqref{eq:sirovichEvpB}. \comment*{SPOD modes}
            \If{$\nwin^{(i)}>3$}{
                $\alpha^{(i)}_{f_k} \gets \abs{(\vb*{\phi}^{(i)}_{1,f_k})^* \vb{W} \vb*{\phi}^{(i-1)}_{1,f_k}}$ \eqref{eq:similarity}. \comment*{similarity}
                \eIf(\comment*[f]{converged}){$1-\alpha^{(i)}_{f_k} \le tol$}{
                    Fix $\vb{\Lambda}_{f_k} \gets \vb{\Lambda}^{(i)}_{f_k}$, $\vb{\Phi}_{f_k} \gets \vb{\Phi}^{(i)}_{f_k}$, and $\nwinf \gets \nwin^{(i)}$.\;
                }(\comment*[f]{not converged}){
                    \uIf{$N_{\mathrm{win},f_{k-1}}$ is fixed {\bf and} $\nwin^{(i)}-N_{\mathrm{win},f_{k-1}}>0$}{
                        \comment{constrained by left neighboring frequency}
                        Fix $\vb{\Lambda}_{f_k} \gets \vb{\Lambda}^{(i)}_{f_k}$, $\vb{\Phi}_{f_k} \gets \vb{\Phi}^{(i)}_{f_k}$, and $\nwinf \gets \nwin^{(i)}$.\;
                    }
                    \ElseIf{$N_{\mathrm{win},f_{k+1}}$ is fixed {\bf and} $\nwin^{(i)}-N_{\mathrm{win},f_{k+1}}>0$}{
                        \comment{constrained by right neighboring frequency}
                        Fix $\vb{\Lambda}_{f_k} \gets \vb{\Lambda}^{(i)}_{f_k}$, $\vb{\Phi}_{f_k} \gets \vb{\Phi}^{(i)}_{f_k}$, and $\nwinf \gets \nwin^{(i)}$.\;
                    }
                }
            }
        }
    }
    \caption{Adaptive SPOD}\label{alg:adapt}
\end{algorithm}
Our proposed adaptive SPOD algorithm is described in more detail in Algorithm~\ref{alg:adapt}, in which the relevant equation numbers are given in parentheses. For simplicity, Algorithm~\ref{alg:adapt} neglects the lossless compression we introduced in Sect.~\ref{sec:compression}, but can be readily modified to include it. Figure~\ref{fig:cavity_nwin_iter} depicts an example of the taper numbers iteratively generated during the adaptive SPOD of the cavity data. The algorithm is initialized at iteration $i=1$ with three tapers for all frequencies. Each subsequent iteration adds one taper to all or part of the spectrum. As a result, the taper numbers and iteration counts are offset by two. Each of the gray contours represents the tapers accumulated during the preceding 50 iterations. Where $\nwin$ plateaus, it has yet to be fixed, and the corresponding leading SPOD mode is not converged. For comparison, the PSD, which following Eq.~\eqref{eq:parseval} is the sum of all eigenvalues at each frequency, is superimposed. The general trend displayed by the contours is that $\nwin$ becomes fixed faster at lower frequencies than at higher frequencies. This trend is a consequence of low-rankness, which leads to rapid modal convergence because it indicates specific physical mechanisms are dominant. In the cavity flow, the low frequencies exhibit low-rankness, whereas the high frequencies consist mainly of broadband turbulence. The former thus converge more quickly. This distinction becomes particularly clear at the second, third, and fourth Rossiter frequencies, denoted by $f_{R_2}$, $f_{R_3}$, and $f_{R_4}$ respectively, and marked by dotted cyan lines. These discrete tones in the PSD coincide with prominent dips in $\nwin$, and are the first frequencies to achieve convergence. We will examine this phenomenon in more detail in Sect.~\ref{sec:cavity}.

\section{Application to mixed tonal and broadband flows: example of an open cavity}\label{sec:cavity}
We first demonstrate sine-taper SPOD on the turbulent open cavity flow from \cite{ZhangEtAl2020ExpFluids}. The cavity flow has a Mach number of $M=u_\infty/c_\infty=0.6$, where $u_\infty$ and $c_\infty$ are the streamwise velocity and speed of sound in the freestream. TR-PIV was used to capture the streamwise and wall-normal velocities, $u$ and $v$, at a time interval of $\Delta t=6.25\times10^{-5}$ {s}, for a total of $N_t=16000$ snapshots. The velocities have been non-dimensionalized by $u_\infty$, {and} lengths by the cavity depth, $D_c$. {For consistency and ease of comparison with \cite{ZhangEtAl2020ExpFluids}, frequencies for the cavity example are reported in units of Hz.} The field of view is resolved by $N_x\times N_y=156\times55$ points. Prior to carrying out the SPOD, the velocity components are arranged in the state vector $\vb{q}=[\vb{u},\vb{v}]^\mathrm{T}$. The weight matrix is the identity, that is, $\vb{W}=\vb{I}$. Since $\vb{q}$ contains the two Cartesian velocity components and the grid is uniform, the norm corresponds to the incompressible turbulent kinetic energy. These parameters are summarized in Table~\ref{tab:dataoverview}.

Instantaneous streamwise velocity fluctuations, $u'$, of the cavity are visualized in Fig.~\ref{fig:dataoverview}a in the introduction. The temporal mean field is shown in Fig.~\ref{fig:dataoverview}c. The fluctuations of the cavity display structures that span a wide range of spatial scales, which is characteristic of broadband turbulent flows. It is well-known that open cavity flows also contain tones associated with the Rossiter mechanism (see \cite{RowleyWilliams2006AnnuRev} for a review). Since we designed the adaptive version of sine-taper SPOD specifically for spectra with mixed tonal and broadband content, in the following we will focus primarily on the cavity data. For non-tonal flows, on the other hand, adaptivity is not necessary. For the broadband jet data in Sect.~\ref{sec:jet} we will thus demonstrate the non-adaptive version of sine-taper SPOD.


\subsection{Preamble: non-adaptive sine-taper SPOD with constant $\nwin$}\label{sec:cavityConstNwin}
\begin{figure}
    \centering
    \includegraphics{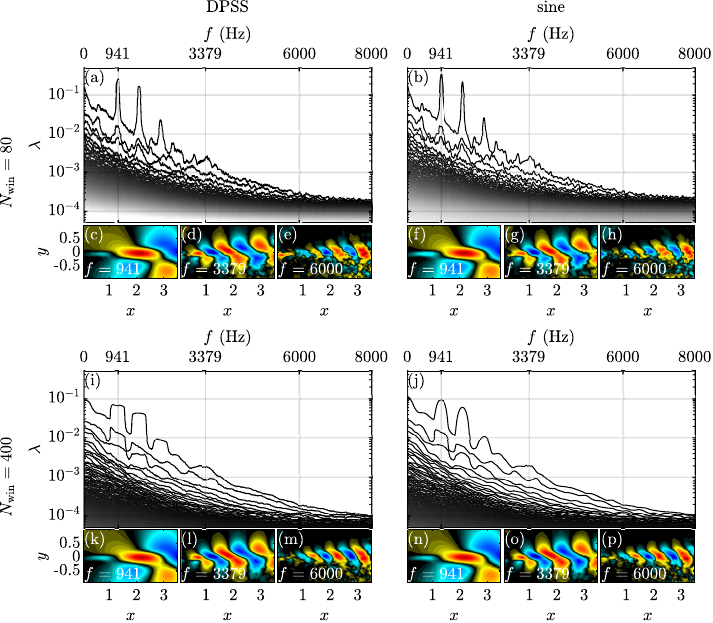}
    \caption{SPOD eigenvalue spectra and leading modes of the cavity, computed with: (a,c,d,e,i,k,l,m) DPSS; (b,f,g,h,j,n,o,p) sine tapers. 80 (a--h) and 400 (i--p) tapers are used; for DPSS, these correspond to setting $\delta f\nfft=40.5$ and 200.5, respectively. The real part of the streamwise velocity component of each mode, $\abs{\vb*{\phi}_u}$, is shown. The colors saturate at $\abs{\vb*{\phi}_u}/\max\abs{\vb*{\phi}_u}=\pm 1$}
    \label{fig:cmp_dpss_sine_spod_spec}
\end{figure}
We begin with a comparison between the state-of-the-art DPSS-taper SPOD and our proposed sine-taper SPOD using a frequency-independent $\nwin$ in the left and right columns of Fig.~\ref{fig:cmp_dpss_sine_spod_spec} respectively. In the standard implementation of DPSS-taper SPOD, Fourier realizations receive uniform weights, $\mu=1/\nwin$. The nonuniform weights in Eq.~\ref{eq:sinParabWgt} are used for sine tapers. The data are not segmented, i.e., $\nfft=N_t=16000$. This demands a potentially large $\nwin$ for convergence. We reiterate that the need for segmenting is eliminated by the computational savings brought about by sine tapers. As Appendix~\ref{sec:appScaling} exemplifies, the memory consumption of sine-taper SPOD is independent of $\nwin$, whereas that of the DPSS version grows linearly with it. This makes DPSS-taper SPOD without segmenting inconvenient for the small cavity data, but prohibitively expensive for the large jet data in Sect.~\ref{sec:jet}.

The eigenvalue spectra in Fig.~\ref{fig:cmp_dpss_sine_spod_spec}a, b, i, and j fade from black to white with increasing mode number. Each row shares the same $\nwin$, giving both taper families approximately equal bandwidths. In Fig.~\ref{fig:cmp_dpss_sine_spod_spec}a--h, $\nwin=80$. Owing to their matching bandwidths, the eigenvalues in Fig.~\ref{fig:cmp_dpss_sine_spod_spec}a and b display similar spectral shapes. In particular, the discrete peaks corresponding to Rossiter tones are precisely localized in frequency, and are visible in at least the three leading eigenvalues. The two spectra possess similar energies in the broadband region, $f\gtrsim4000$ {Hz}. However, the sine-tapered spectrum in Fig.~\ref{fig:cmp_dpss_sine_spod_spec}b is distinguished by sharper peaks and higher energies at the tonal frequencies, e.g., $f=941$ {Hz}, which indicates lower local bias. Bin-to-bin variance, which can obscure important features, especially in shorter data, affects both spectra but is more severe in the DPSS-tapered spectrum. Below each spectrum, representative modes at three frequencies, $f=941$, 3379, and 6000 {Hz}, are shown. Each mode, or eigenvector, is determined only up to an arbitrary phase. Without loss of generality, we approximately align the phases of the modes at the same frequency to faciliate comparison. At each frequency, DPSS- and sine-tapered modes are visually nearly indistinguishable. The modes at the low frequency, $f=941$ {Hz}, a Rossiter tone, have very smooth waveforms that indicate they are fully converged. This is expected since the underlying Rossiter instability ensures the first and second eigenvalues, $\lambda_1$ and $\lambda_2$, are separated by a large gap in energy.\footnote{A large separation between $\lambda_1$ and $\lambda_2$ is sufficient but not necessary for rapid modal convergence. A counterexample is a flow in which two different physical mechanisms are active at the same frequency, leading to a large separation between $\lambda_2$ and $\lambda_3$ instead. This may occur if, e.g., statistical symmetries are present. An example can be found in \cite{YeungEtAl2022AIAA}.} For DPSS, $\lambda_1-\lambda_2\approx0.25$; for sine, $\lambda_1-\lambda_2\approx0.32$. Such low-rank behavior suggests that the structures are prevalent and deterministic, therefore rapidly converged. The leading eigenvalue at the intermediate frequency, $f=3379$ {Hz}, also reveals a peak. However, at this frequency, $\lambda_1-\lambda_2\approx0.0017$ and 0.0016 for DPSS and sine tapers, respectively. Consistent with the smaller eigenvalue separation, the modes are less well-converged, though they continue to display the clear streamwise and wall-normal wavenumbers that are characteristic of Kelvin-Helmholtz (KH) wavepackets. At the high frequency, $f=6000$ {Hz}, the two leading eigenvalues possess similar energies. The mode at this frequency exemplifies the slow rate of convergence in the broadband region. The KH wavepackets, while visible, appear distorted and lack a clearly delineated envelope. Overall, these results are consistent with the spectrum and modes obtained by \cite{ZhangEtAl2020ExpFluids} using the Welch method.

If $\nwin$ is increased to 400, the leading modes at the Rossiter frequency, $f=941$ {Hz}, in Fig.~\ref{fig:cmp_dpss_sine_spod_spec}k and n remain visually unchanged, since they are already converged with $\nwin=80$. The modes at the intermediate and high frequencies, $f=3379$ and 6000 {Hz}, on the other hand, have become smoother and more compact, though their convergence may be further improved with an even greater $\nwin$. However, raising $\nwin$ further would be inadvisable, as the obvious and undesired broadening of the tones in the spectra in Fig.~\ref{fig:cmp_dpss_sine_spod_spec}i and j demonstrate. As expected, with $\nwin=400$ the leading eigenvalues of both spectra have lower energy at all frequencies in comparison with $\nwin=80$. This phenomenon is related to the energy being distributed over a much higher number of modes. The sine-tapered spectrum in Fig.~\ref{fig:cmp_dpss_sine_spod_spec}j continues to preserve more energy at the tonal frequencies than the DPSS-tapered spectrum does in Fig.~\ref{fig:cmp_dpss_sine_spod_spec}i. In both spectra, however, the peaks are drastically flattened as the larger $\nwin$ effectively averages over a larger bandwidth. Less energetic, closely-spaced peaks are merged. At the peaks, the DPSS spectrum takes on a characteristic boxy shape \cite{PrietoEtAl2007GJI}, whereas the sine spectrum retains higher curvature. This means that despite the excessive broadening, the frequencies of the most dominant Rossiter tones can still be accurately estimated from the sine spectrum. In the DPSS spectrum, the tonal frequencies are much less distinct, and have to be estimated as the midpoint of each of the flattened peaks. From these observations we draw two conclusions: (i) DPSS- and sine-taper SPOD demonstrate similar variance reduction, with a slight advantage to sine tapers; however, the latter excels at resolving tones; and (ii) in the broadband region a higher $\nwin$ is desirable because it leads to lower variance and more converged modes; at tonal frequencies, however, it is not, since well-converged modes and accurately-resolved peaks are obtained with a low $\nwin$. These conclusions motivate the use of the adaptive algorithm for mixed tonal and broadband flows.

\begin{figure}
    \centering
    \includegraphics{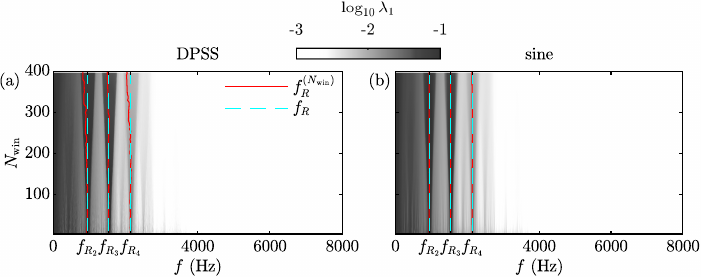}
    \caption{Leading SPOD eigenvalues of the cavity computed with: (a) DPSS; (b) sine tapers. The solid red lines correspond to the peak frequencies of the three most prominent Rossiter tones for each $\nwin$. For comparison, the dashed cyan lines mark the actual tonal frequencies, $f_{R_2}$, $f_{R_3}$, and $f_{R_4}$, previously determined in Fig.~\ref{fig:cavity_nwin_iter}. (a,b) share the same contour levels}
    \label{fig:cmp_dpss_sine_spodcontour}
\end{figure}
Next we systematically and quantitatively investigate the effect of $\nwin$ on tonal frequency estimation. The contours in Fig.~\ref{fig:cmp_dpss_sine_spodcontour} show the evolution of the leading SPOD eigenvalues as $\nwin$ is increased from five to 400 for DPSS (\ref{fig:cmp_dpss_sine_spodcontour}a) and sine (\ref{fig:cmp_dpss_sine_spodcontour}b) tapers. Darker shades of gray represent larger eigenvalues. In both spectra, thin striations in the region $\nwin\lesssim50$ indicate high bin-to-bin variance. For higher $\nwin$, the striations are smoothed and replaced by bands that gradually widen with bandwidth. Dark bands centered at $f_{R_2}$, $f_{R_3}$, and $f_{R_4}$ correspond to the Rossiter frequencies. Accurately locating these frequencies becomes more challenging as the bands are broadened, especially with DPSS tapers. As Fig.~\ref{fig:cmp_dpss_sine_spod_spec}i showed, when $\nwin$ is high, DPSS tapers flatten peaks, making the peak frequencies ambiguous. In Fig.~\ref{fig:cmp_dpss_sine_spodcontour}, for each $\nwin$ we overlay the peak frequency within each dark band,
\beq
f^{(\nwin)}_R = \underset{|f_k-f_R|<\frac{\delta f}{\Delta t}}{\arg\max}\lambda^{(\nwin)}_{1,f_k},
\eeq
as solid red lines. The dashed cyan lines mark the actual Rossiter frequencies determined in Fig.~\ref{fig:cavity_nwin_iter}. In Fig.~\ref{fig:cmp_dpss_sine_spodcontour}a, for $\nwin\lesssim200$, the peak frequencies $f^{(\nwin)}_R$ of the DPSS spectrum initially stay close to---but meander around---the reference frequencies $f_R$. The meandering is a consequence of high bin-to-bin variance. For $\nwin\gtrsim200$, $f^{(\nwin)}_R$ becomes biased towards low frequencies. The leftward bias grows with $\nwin$, as a larger bandwidth increasingly merges the discrete tones with the underlying broadband spectrum, which decays with frequency. Although the meandering and frequency bias are both observed in the sine spectrum in Fig.~\ref{fig:cmp_dpss_sine_spodcontour}b, they are significantly subtler here than they are in Fig.~\ref{fig:cmp_dpss_sine_spodcontour}a. The tones in Fig.~\ref{fig:cmp_dpss_sine_spodcontour}b are tracked smoothly and more accurately for all $\nwin$, achieving variance reduction without compromising tonal frequency estimation. These observations apply to peaks that are distant from each other in frequency, but are likely to break down otherwise. If neighboring peaks merge, as some have in Fig.~\ref{fig:cmp_dpss_sine_spod_spec}j for $\nwin=400$, their frequencies will not be resolved even by sine tapers. Nonetheless, the favorable performance of sine-taper SPOD demonstrates that minimizing local bias should be prioritized over maximizing spectral concentration when the spectrum comprises both discrete and broadband components.

\begin{figure}
    \centering
    \includegraphics{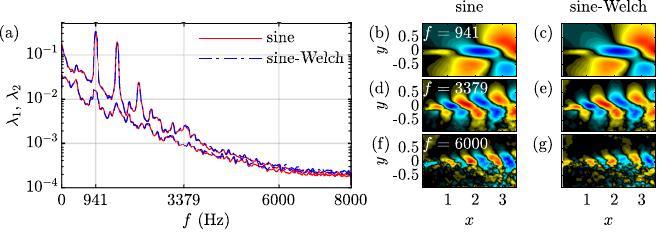}
    \caption{First two SPOD eigenvalues, $\lambda_1$ and $\lambda_2$, and leading modes of the cavity at $f=941$ (b,c), 3379 (d,e), and 6000 (f,g) {Hz}. The solid red spectrum in (a) and modes in (b,d,f) use multitaper estimation, with a block size of $\nfft=N_t$ and 97 sine tapers. The dash-dotted blue spectrum in (a) and modes in (c,e,g) use multitaper-Welch estimation, with a block size of $\nfft=4096$, 50\% overlap, and 24 sine tapers}
    \label{fig:cmp_sinWelch_sin_spod}
\end{figure}
When memory requirements become prohibitive for long data records, \cite{Schmidt2022TCFD} suggested using a hybrid multitaper-Welch SPOD, combining segmenting with multitaper SPOD. As discussed in Sect.~\ref{sec:multitaper}, sine-taper SPOD with compression diminishes the need for segmenting. Figure~\ref{fig:cmp_sinWelch_sin_spod} compares the sine-taper and sine-Welch approaches. For sine-taper SPOD, the full-length data are windowed using 97 sine tapers. For sine-Welch SPOD, the snapshots are segmented into six blocks of length $\nfft=4096$ each, with 50\% overlap between adjacent blocks. Each block is then windowed using 24 sine tapers. These sets of parameters result in similar bandwidths, $\delta f$, which in turn produce spectral estimates with comparable variance and bias. The leading two eigenvalues, $\lambda_1$ and $\lambda_2$, in Fig.~\ref{fig:cmp_sinWelch_sin_spod}a and leading modes in Fig.~\ref{fig:cmp_sinWelch_sin_spod}b--g produced by the sine-taper and sine-Welch methods match each other closely. The consistency between the two approaches shows that either allows for good variance and bias control, provided that spectral estimation parameters are chosen judiciously. For the sine-Welch estimator, $\nfft$ and overlap must be selected in addition to $\nwin$; for the sine estimator, only $\nwin$ needs to be selected. Its maximal frequency resolution and small number of parameters make sine-taper SPOD without segmenting advantageous in most cases, in particular for the proposed adaptive algorithm.

\subsection{Adaptive sine-taper SPOD}\label{sec:cavityAdaptNwin}
\begin{figure}
    \centering
    \includegraphics{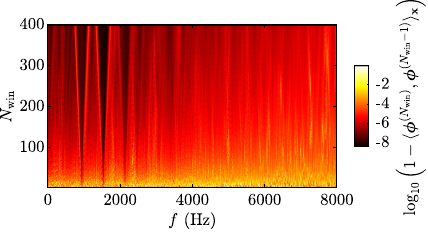}
    \caption{Similarity between the leading SPOD mode of the cavity at each frequency estimated using $\nwin$ sine tapers and the corresponding mode estimated using $\nwin-1$ tapers, for $\nfft=N_t$. Dark red indicates high mode similarity}
    \label{fig:spod_modesim}
\end{figure}
As previously seen in Fig. 3, while sine-taper SPOD improves upon the existing DPSS-taper algorithm, for the cavity flow it remains unable to provide satisfactory spectral and modal estimates at all frequencies using a single $\nwin$. Recall that in the right column of Fig.~\ref{fig:cmp_dpss_sine_spod_spec}, $\nwin=80$ generates a spectrum with precisely localized tones but high bin-to-bin variance, and poorly converged modes at high frequencies. A fivefold increase in $\nwin$ reduces variance and enhances convergence at the expense of unacceptable local bias. Adaptive sine-taper SPOD, introduced in Sect.~\ref{sec:adaptive}, grants greater control of the variance-bias trade-off by allowing $\nwin$ to vary with frequency. It determines a frequency-local $\nwinf$ for each frequency based on the rate of convergence of the leading mode. The contours in Fig.~\ref{fig:spod_modesim} display the similarity, $\alpha^{(\nwin)}_{f_k}$, defined in Eq.~\eqref{eq:similarity}, between the leading modes obtained using $\nwin$ and $\nwin-1$ tapers. As each mode converges, $\alpha\to1$ and $\log_{10}(1-\alpha)\to-\infty$, producing darker contours. Overall the contours darken with increasing $\nwin$, confirming that convergence improves with the addition of tapers. Meanwhile, the contours lighten with increasing frequency in the region $f\gtrsim3000$ {Hz} characterized by broadband turbulence. The darkest wedges, where $\alpha$ is closest to one, are centered around the Rossiter frequencies. The fastest converging frequencies are thus the ones which exhibit low-rank behavior in Fig.~\ref{fig:cmp_dpss_sine_spod_spec}; conversely, convergence slows at high frequencies, where the leading and suboptimal eigenvalues cluster around each other. The bin-to-bin variance of the similarity, $\alpha$, underscores the need to enforce constraint~\eqref{eq:nwinConstrain} on $\nwinf$ explicitly.

\begin{figure}
    \centering
    \includegraphics{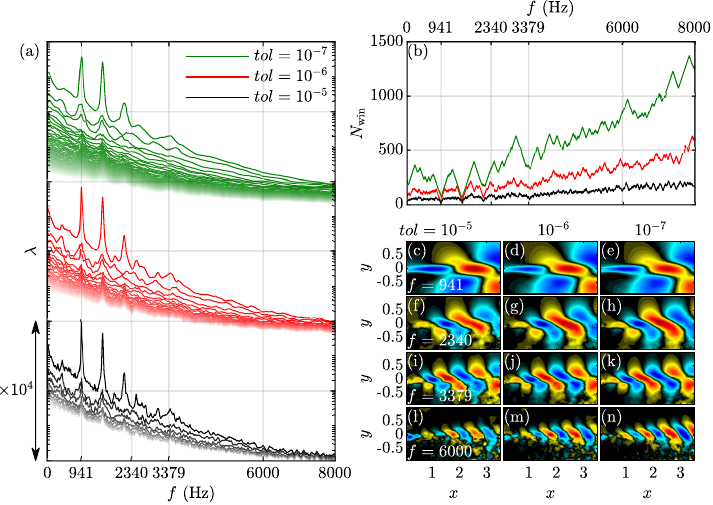}
    \caption{Adaptive SPOD eigenvalue spectra of the cavity for different similarity tolerances (a), the corresponding frequency-local taper numbers (b), and the leading modes at $f=941$ (c--e), 2340 (f--h), 3379 (i--k), and 6000 (l--n) {Hz}, for $tol=10^{-5}$ (black), $10^{-6}$ (red), and $10^{-7}$ (green). For ease of comparison, the red and green spectra have been shifted. The first $\min(\nwinf)$ adaptive SPOD eigenvalues are shown in (a)}
    \label{fig:spod_tap2tap}
\end{figure}
To study the effects of the similarity tolerance, $tol$, Fig.~\ref{fig:spod_tap2tap} shows the adaptive SPOD eigenvalue spectra, leading modes, and taper numbers of the cavity flow for $tol=10^{-5}$, $10^{-6}$, and $10^{-7}$. For each spectrum, at each frequency only the first $\min(\nwinf)$ eigenvalues out of a total of $\nwinf$ are displayed. For $tol=10^{-5}$, the Rossiter tones are sharp and well-localized in frequency, and the bin-to-bin variance is high. The taper number fluctuates between approximately 10 and 200. The Rossiter tones coincide with local minima in $\nwinf$. In particular, at the second and third Rossiter frequencies, $f=941$ and 1538 {Hz}, $\nwinf\approx10$. Among the modes in Fig.~\ref{fig:spod_tap2tap}c, f, i, and l, only the one at $f=941$ {Hz} is smooth. The mode at $f=6000$ {Hz} contains KH wavepackets that appear disjointed and distorted. When we tighten the tolerance to $10^{-6}$, the bin-to-bin variance of the spectrum is greatly reduced at all frequencies. Unlike in Fig.~\ref{fig:cmp_dpss_sine_spod_spec}j, however, here the tones remain sharp and individually resolved. This low local bias is enabled by large variations in $\nwinf$. While $\nwinf$ is generally much higher for $tol=10^{-6}$ than for $10^{-5}$, it dips to around 30 at the most prominent tones. The modes have also become smoother. This is most apparent at $f=6000$ {Hz}, where the wavepackets acquire clearly-defined streamwise and wall-normal wavenumbers. Reducing the tolerance to $10^{-7}$ necessitates still higher $\nwinf$. The bin-to-bin variance of the spectrum is now largely eliminated at the expense of more severe leakage. Some of the closely-spaced peaks have merged with their neighbors, though the Rossiter tones remain sharp and their broadening is limited. The modes again show smoother waveforms, free of distortion or artifacts. 

Of the three similarity tolerances, we deem $tol=10^{-6}$ to result in both low bias and low variance. From Fig.~\ref{fig:cmp_dpss_sine_spod_spec} we concluded that with a frequency-independent $\nwin$, the variance-bias trade-off cannot be adequately controlled in all regions of a tonal spectrum. Compared to non-adaptive SPOD, the adaptive estimator combines the low bias of a small $\nwin$ to resolve discrete tones with the low variance of a large $\nwin$ to estimate the broadband portions of the spectrum. The tones in the $tol=10^{-6}$ spectrum in Fig.~\ref{fig:spod_tap2tap}a experience a similar degree of broadening to those in the $\nwin=80$ spectrum in Fig.~\ref{fig:cmp_dpss_sine_spod_spec}b, though the former attain higher energies. The bin-to-bin variance of the $tol=10^{-6}$ spectrum, on the other hand, is nearer that of the $\nwin=400$ spectrum in Fig.~\ref{fig:cmp_dpss_sine_spod_spec}j. The modes computed using $tol=10^{-6}$ in Fig.~\ref{fig:spod_tap2tap}d, j, and m resemble those using $\nwin=400$ in Fig.~\ref{fig:cmp_dpss_sine_spod_spec}n, o, and p respectively. In summary, with the right similarity tolerance, the proposed adaptive algorithm simultaneously achieves accurate resolution of tonal peaks and smooth estimation of broadband spectra.


\begin{figure}
    \centering
    \includegraphics{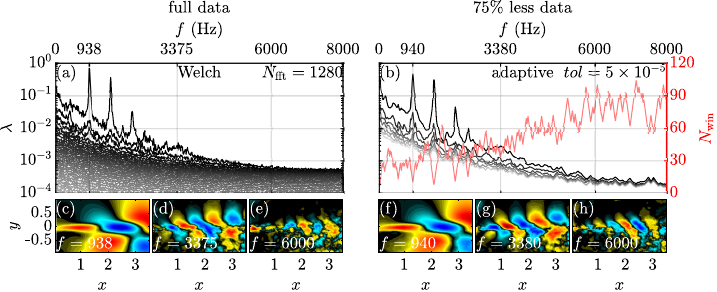}
    \caption{SPOD eigenvalues and leading modes of the cavity computed from: (a,c--e) all 16000 snapshots using the Welch method; (b,f--h) the first 4000 snapshots, or 75\% less data, using adaptive sine-taper SPOD. The Welch SPOD uses $\nfft=1280$ with 75\% overlap and the Hamming window---the same parameters as in \cite{ZhangEtAl2020ExpFluids}. The adaptive SPOD uses $tol=5\times10^{-5}$. For the adaptive case, $\nwin$ is overlaid}
    \label{fig:cavity_spod_025data}
\end{figure}
Next we demonstrate that adaptive SPOD only requires a fraction of the data to produce estimates comparable to those of standard Welch SPOD. Figure~\ref{fig:cavity_spod_025data}a shows the Welch SPOD spectrum obtained by \cite{ZhangEtAl2020ExpFluids}, the work in which the cavity data were originally published, using the full data, with parameters $\nfft=1280$, 75\% overlap, and the Hamming window. In contrast, the adaptive SPOD spectrum shown in Fig.~\ref{fig:cavity_spod_025data}b uses only the first 4000 snapshots, i.e., 25\% of the original data. The similarity tolerance is adjusted such that the tones in both spectra are approximately equally resolved. We have confirmed that the differences in magnitudes of the tones between the two spectra are associated with the reduced test data segment, not with the estimators. The adaptive spectrum exhibits less bin-to-bin variance than the Welch spectrum does. The modes at $f\approx940$ {Hz} in Fig.~\ref{fig:cavity_spod_025data}c and f are comparable, whereas those at $f\approx3380$ and 6000 {Hz} in Fig.~\ref{fig:cavity_spod_025data}d, e, g, and h appear slightly smoother and more distinct with the adaptive estimator. As Fig.~\ref{fig:spod_tap2tap} shows, given more data adaptive SPOD will perform even better. Nevertheless, its efficient usage of scarce data is a major advantage over standard SPOD.

\section{Application to broadband flows: example of a turbulent jet}\label{sec:jet}
\begin{figure}
    \centering
    \includegraphics[width=\linewidth]{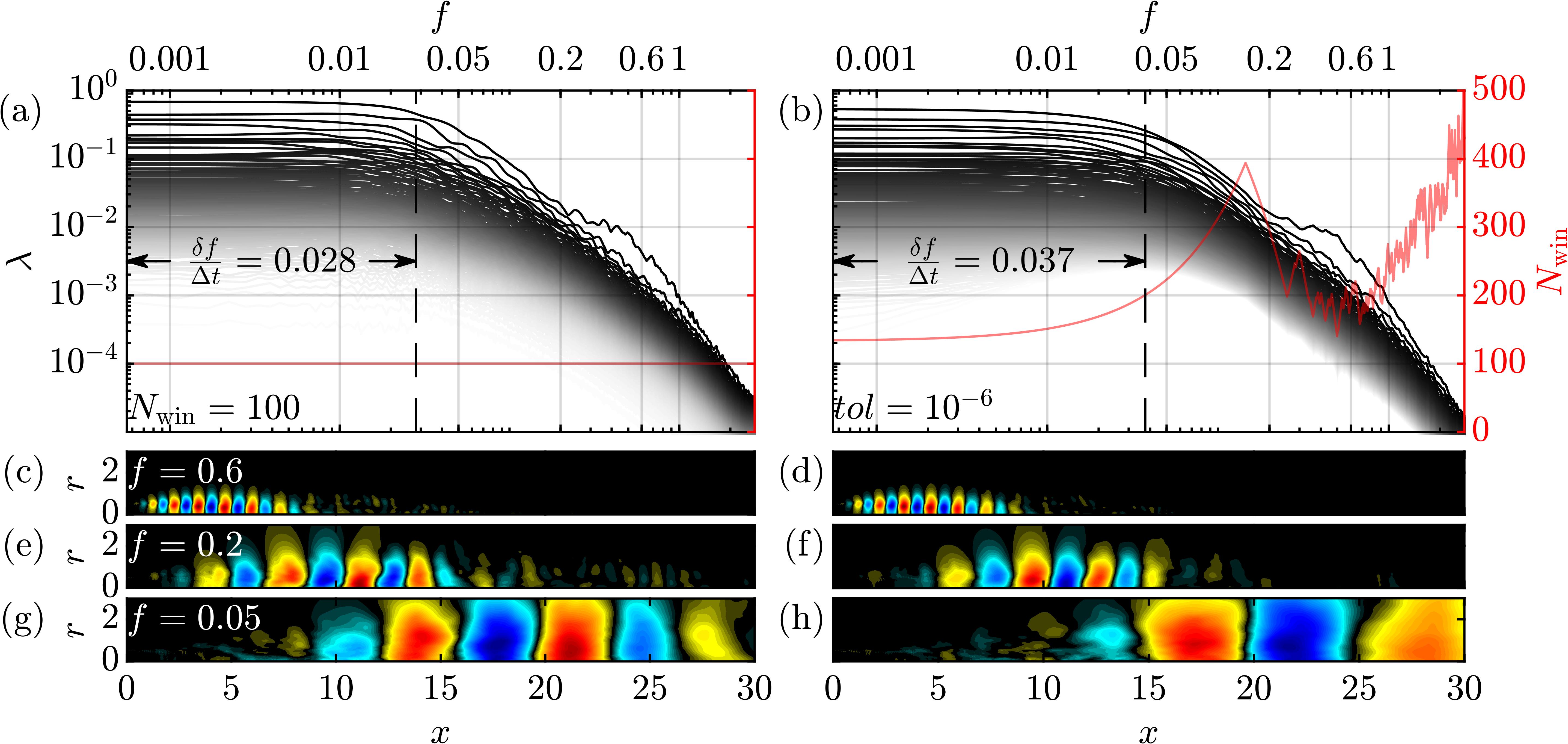}
    \caption{SPOD eigenvalues (black) and $\nwin$ (red) of the jet flow computed using: (a) $\nwin=100$; (b) $tol=10^{-6}$, along with their respective leading modes at frequencies $f=0.6$ (c,d), 0.2 (e,f), and 0.05 (g,h). In (b), only the first $\min\nwinf$ eigenvalues are shown. Dashed lines mark the half-bandwidth, $\delta f/\Delta t$, of the $f=0$ bin in (a,b)}
    \label{fig:jet_spod}
\end{figure}
Having demonstrated sine-taper SPOD on the open cavity flow, we next examine its application to the round jet flow from \cite{BresEtAl2018JFM}, first studied using standard Welch SPOD by \cite{SchmidtEtAl2018JFM}. The jet is an example of a broadband turbulent flow, for which we explicitly highlight that the adaptive algorithm is not recommended. Instead we recommend using a constant number of sine tapers to benefit from the increased frequency resolution. The jet flow has a Mach number of $M=u_j/c_j=0.9$, where $u_j$ and $c_j$ are the centerline velocity and sound speed at the nozzle exit, respectively. Large-eddy simulation (LES) using the solver `Charles' \cite{BresEtAl2017AIAAJ} produced $N_t=10000$ snapshots at an interval of $\Delta t=0.18$. Each snapshot contains the density, $\rho$, axial, radial, and azimuthal velocities, $u_x$, $u_r$, and $u_\theta$, and temperature, $T$. The density, velocities, and temperature have been non-dimensionalized by their values on the centerline at the nozzle exit, $\rho_j$, $u_j$, and $T_j$, respectively. Lengths are non-dimensionalized by the nozzle diameter, $D_j$, and time by $D_j/u_j$. {Frequencies for the jet example are thus expressed as Strouhal numbers, $fD_j/u_j$. For brevity, we denote them by $f$, which should be understood to be dimensionless.} The unstructured LES data are interpolated onto a nonuniform cylindrical grid of size $N_x\times N_r\times N_\theta=656\times138\times128$. The rotational symmetry of the nozzle permits independent consideration of each azimuthal wavenumber component. Only the axisymmetric component, with wavenumber $m=0$, will be examined in this work. The five variables are assembled in the state vector $\vb{q}=[\hat{\vb*{\rho}},\hat{\vb{u}}_x, \hat{\vb{u}}_r, \hat{\vb{u}}_\theta, \hat{\vb{T}}]^\mathrm{T}$. The weight matrix $\vb{W}$ discretizes the inner product 
\beq
\expval{\vb*{q}_1,\vb*{q}_2}_{\vb*{x}}=\iiint_\Omega \vb*{q}^*_2 \mathrm{diag}\qty(\qty[\frac{\bar{T}}{\gamma \bar{\rho} M^2},\bar{\rho},\bar{\rho},\bar{\rho},\frac{\bar{\rho}}{\gamma(\gamma-1)\bar{T}M^2}]) \vb*{q}_1 r\dd{x}\dd{r}\dd{\theta},
\eeq
where $\Omega$ denotes the domain and $\gamma$ is the adiabatic constant. Thus the eigenvalues are the volume-integrated compressible energy \cite{Chu1965ActaMech} of each mode. The parameters are summarized in Table~\ref{tab:dataoverview}. We note that the higher spatial resolution and higher number of variables of the jet database make it an order of magnitude larger than the cavity database.

\begin{figure}
    \centering
    \includegraphics{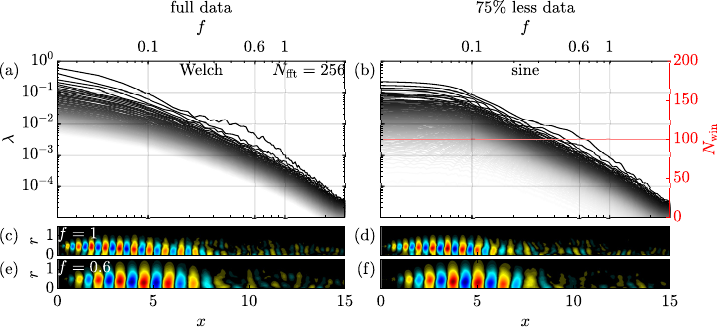}
    \caption{SPOD eigenvalues and leading modes of the jet computed from: (a,c,e) all 10000 snapshots using the Welch method; (b,d,f) the first 2500 snapshots, or 75\% less data, using non-adaptive sine-taper SPOD. The Welch SPOD uses $\nfft=256$ with 50\% overlap and the Hamming window---the same parameters as in \cite{SchmidtEtAl2018JFM}. The sine-taper SPOD uses $\nwin=100$}
    \label{fig:jet_spod_025data}
\end{figure}

The broadband nature of the SPOD spectrum of the jet \cite{SchmidtEtAl2018JFM} favors a constant rather than adaptive $\nwin$. In general, however, we may not be able to exploit \textit{a priori} knowledge of a particular flow. Without such foresight, distinguishing a tonal spectrum from a noisy estimate of a broadband spectrum may not be a trivial task. Adaptive sine-taper SPOD, while not designed for broadband flows, nevertheless must be robust in such cases. For the jet, Fig.~\ref{fig:jet_spod}b shows the adaptive SPOD eigenvalues and taper numbers computed using $tol=10^{-6}$ and $\nfft=N_t=10000$. Compared to the spectrum produced by \cite{SchmidtEtAl2018JFM} using Welch SPOD {(reproduced in Fig.~\ref{fig:jet_spod_025data}a)}, the adaptive spectrum possesses a similar shape. Notably, the adaptivity in the high-frequency broadband regions does not generate numerical artifacts. Unlike the cavity flow in Fig.~\ref{fig:spod_tap2tap}b, here $\nwinf$ is uninterrupted by deep valleys that signify dramatic reductions in the local bandwidth. Consistent with the absence of tones, the taper number remains moderately high: $\nwinf\gtrsim 100$ over the entire frequency range. $\nwinf$ does experience a spike in the range $0.04\lesssim f \lesssim 0.3$, which is associated with the KH mechanism becoming dominant for $f\gtrsim0.2$ (see \cite{SchmidtEtAl2018JFM}). The slowing of mode convergence when the eigenvalue separation is small is thus common to both the jet and the cavity.

Below $f\approx0.05$, the spectrum gradually flattens into a plateau. The half-width of this plateau is comparable to the half-bandwidths, $\delta f/\Delta t$, of the low frequencies. As an example, the half-bandwidth of the $f=0$ bin, $\delta f/\Delta t=0.037$, is indicated in Fig.~\ref{fig:jet_spod}. The shape of the plateau is due to the main lobe of the mean spectral window (i.e., mean PSD) of DPSS and sine tapers \cite{Thomson1982IEEE,RiedelSidorenko1995IEEE,WaldenEtAl1995Biomet}, which becomes nearly rectangular with width equal to the bandwidth as $\nwin$ grows. In effect, the plateau---and therefore the bandwidth---determines the lowest nonzero frequency resolvable by the spectrum.

In Fig.~\ref{fig:jet_spod}a, the spectrum windowed using $\nwin=100$ tapers is shown for reference, and is consistent with the adaptive spectrum in Fig.~\ref{fig:jet_spod}b. The leading modes at $f=0.6$ in Fig.~\ref{fig:jet_spod}c and d, as well as those at $f=0.2$ in Fig.~\ref{fig:jet_spod}e and f, are also well-matched, though the adaptively tapered mode at $f=0.2$ has a smoother, more compact waveform. At $f=0.05$, however, the adaptive mode in Fig.~\ref{fig:jet_spod}h significantly underestimates the wavenumber, $k$, of the structures. This discrepancy occurs only for $0.04\lesssim f \lesssim 0.2$, where $\nwinf$ spikes, and suggests that in the low-frequency region averaging over a large bandwidth may bias modes towards lower wavenumbers, or equivalently lower frequencies. This phenomenon will be discussed in Sect.~\ref{sec:discussion}. We conclude that the $\nwin=100$ estimate is better, and recommend non-adaptive sine-taper SPOD for broadband flows to take advantage of the high frequency resolution of the method. An $\nwin$ of 100 is a good starting point, but should be adjusted to balance variance and bias.

Analogous to Fig.~\ref{fig:cavity_spod_025data}, Fig.~\ref{fig:jet_spod_025data} compares the Welch SPOD obtained by \cite{SchmidtEtAl2018JFM} using the full jet data, and non-adaptive sine-taper SPOD with 75\% of the data discarded. For the latter, 100 tapers are used. The sine-taper spectrum displays lower variance than the Welch spectrum does. Both algorithms produce modes of comparable smoothness. We conclude that the non-adaptive sine-taper algorithm also requires a fraction of the snapshots that the standard Welch algorithm demands for broadband flows, just as the adaptive version does for tonal flows.

\section{Discussion}\label{sec:discussion}
Distinct from spectral estimation of one-dimensional signals, SPOD requires the estimation of both modes and their corresponding energies. The latter, in particular, must be well-converged to aid in physical discovery and reduced-order modeling efforts. An accurate estimate of SPOD modes and modal energies hinges on satisfying two competing objectives: {on one hand, minimizing the variance of the modes and energies, and on the other, minimizing their bias}. With standard Welch-SPOD, the user attempts to balance both objectives by selecting an $\nfft$. The recently-introduced multitaper SPOD of \cite{Schmidt2022TCFD}, which uses the family of DPSS tapers, affords finer-grained control of this balance by adding $\nwin$ as an input. Its shortcomings include excessive bias when estimating tonal spectra, and high computational cost. In this work, we introduce sine tapers to multitaper SPOD. By construction, sine tapers minimize local bias. They also do not require the explicit computation and storage of $\nwin$ Fourier transforms. Instead, sine-tapered Fourier realizations can be recovered from the central differences between pairs of Fourier modes sampled from a single FFT. In addition, sine tapers enable $\nwin$ to vary with frequency and adapt to rapid variations of the spectrum within a given bandwidth, making sine-taper SPOD well-suited to tonal spectra.

To further protect tones in the spectrum from bias, we propose an iterative procedure that uses modal convergence to determine the taper number, $\nwinf$, to apply to each frequency. Starting from a small taper number, we incrementally add tapers to all frequencies, but fix $\nwinf$ for any frequencies at which the leading SPOD mode has converged to tolerance. For the remaining frequencies, the process is repeated until all leading modes are converged. The frequencies that converge faster tend to be characterized by tones or other low-rank dynamics. As a result, the bandwidths at these frequencies become narrower than those in broadband regions. The similarity tolerance is the only parameter that the user need select. For large data, SPOD, especially multitaper SPOD, may become intractable. To further alleviate high computational complexity and memory usage, we losslessly compress the data in the time domain via a QR or eigenvalue decomposition. To a large extent, the combination of sine tapers and lossless compression frees the user from limitations on estimation parameters due to a lack of local computational resources. For the TR-PIV data of the cavity flow, compared to DPSS-taper SPOD with a constant $\nwin$, we find adaptive sine-taper SPOD enhances the resolvability of tones, reduces bin-to-bin variance {of the spectrum}, and improves mode convergence. For the same similarity tolerance, we observe similar bin-to-bin variance and mode smoothness for the turbulent jet data. After testing on an additional TR-PIV database \cite{SeoEtAl2023ExpFluids} and a direct numerical simulation (DNS) database (with similar parameters to \cite{MathiasMedeiros2020AIAA}), we feel confident in recommending $tol=10^{-6}$ as best practice. While it is always beneficial to obtain more data, the adaptive SPOD of tonal flows produces results similar to those of standard Welch SPOD with up to 75\% less data. 

In Sect.~\ref{sec:cavityConstNwin} we compared sine-taper SPOD with and without segmenting. While segmenting can provide comparable results, avoiding it offers greater simplicity and maximizes frequency resolution. If the data record is very long, however, allowing $\nfft$ to grow with $N_t$ may become untenable. This is easily prevented by capping $\nfft$ at some fixed value.
Care must also be exercised when interpreting the total power (i.e., area under the curve) of an adaptively tapered spectrum. For the adaptive version of Thomson's multitaper estimator applied to one-dimensional signals \cite{Thomson1982IEEE}, it can be shown \cite{ParkEtAl1987JGR,PercivalWalden1993Cambridge} that Parseval's Theorem in Eq.~\eqref{eq:parseval} is no longer guaranteed to be satisfied, even in expected value. For our adaptive sine-taper SPOD, we also observe this empirically. In practice this effect is modest; the total power of the $tol=10^{-6}$ spectrum of the cavity in Fig.~\ref{fig:spod_tap2tap}a, for instance, exceeds the sample variance only by 2\%. The jet example in Fig.~\ref{fig:jet_spod} highlights a different limitation of multitaper SPOD: when $\nwin$ is large, modes may be biased in wavenumber and frequency. This bias is common to DPSS and sine tapers, but most severe for the adaptive algorithm. It arises due to partial correlation between modes at different frequencies within a given bandwidth. The bias becomes more pronounced at small $f$, since a given bandwidth encompasses structures of more disparate wavenumbers at low frequencies than it does at high frequencies. Because the underlying spectrum decays with frequency, low-frequency structures are energetically dominant, biasing the estimated modes towards $f=0$ and $k=0$. This is another manifestation of the variance-bias trade-off. The bias can be alleviated with a variable $tol$ that decreases with frequency, sacrificing convergence for small $f$. In designing the adaptive algorithm for tonal flows, we have prioritized protecting tones and controlling variance over extracting low-frequency modes. For the most accurate estimation of broadband flows, adaptivity should be switched off entirely. Instead we recommend a constant $\nwin$ to obtain spectral estimates that retain the advantages of non-adaptive sine-taper SPOD over the DPSS and Welch estimators.


\section*{Declaration of Interests}

\noindent{\bf Ethical Approval.} Not applicable.

\noindent{\bf Funding.} BCYY and OTS gratefully acknowledge support from Office of Naval Research grant N00014-23-1-2457, under the supervision of Dr. Steve Martens. The TR-PIV data were created with support from AFOSR Award Number FA9550-17-1-0380. Creation of the LES data was supported by NAVAIR SBIR under the supervision of J. T. Spyropoulos, with computational resources provided by DoD HPCMP at the ERDC DSRC supercomputer facility.  
 
\noindent{\bf Availability of data and materials}. Data is available on request from the authors. Code is openly available in a public repository.

\noindent{\bf Acknowledgements}. We thank Lou Cattafesta and Yang Zhang for providing the cavity TR-PIV data, and Guillaume Br\`es for generating the jet LES data. {We are grateful to the three anonymous referees for their comments.} BCYY thanks Tianyi Chu for suggesting using an EVD for orthogonalization.

\appendix

\section{Wall time and memory usage}\label{sec:appScaling}
\begin{figure}
    \centering
    \includegraphics{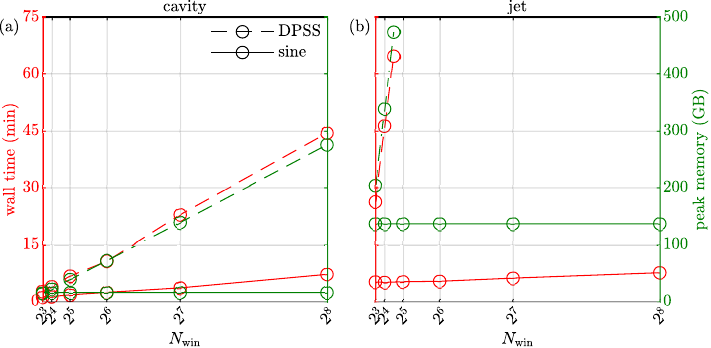}
    \caption{Wall-clock time (red) and peak memory usage (green) of DPSS-taper (- - -) and sine-taper (------) SPOD: (a) cavity; (b) jet}
    \label{fig:scaling}
\end{figure}
Figure~\ref{fig:scaling} reports the wall-clock time and peak memory usage for the cavity and jet data as $\nwin$ is varied, using non-adaptive multitaper SPOD. Tests were conducted using MATLAB on a workstation with 48 Intel Xeon cores and 500 GB of memory. The most time-consuming operations, including QR, FFT, and EVD, were performed in single precision. At each frequency, only the three leading SPOD modes were stored. For DPSS-taper SPOD, the largest $\nwin$ tested for the jet was limited by available memory to 24. For sine-taper SPOD, lossless compression was achieved using QR and EVD for the cavity and jet, respectively.

The wall time and peak memory consumption of DPSS-taper SPOD grow linearly with $\nwin$. For sine-taper SPOD, the wall time is dominated by compression of the snapshots and expansion of the SPOD modes; memory usage is dominated by storage of the snapshots and the modes. As a result, the wall time of sine-taper SPOD depends only weakly on $\nwin$, while its memory usage stays constant. For the same reason, the memory usage of adaptive sine-taper SPOD is comparable to that of the non-adaptive algorithm. The adaptive algorithm is considerably slower, however. With $tol=10^{-6}$, adaptive SPOD of the cavity and jet examples take approximately 18 and eight hours, respectively, to complete.
Although the DPSS-taper and Welch algorithms may also benefit from compression, they require computing and storing an FFT for every taper; the sine-taper algorithm, on the other hand, uses the central difference identity in Eq.~\eqref{eq:ctrDiff}, thus requiring only a single FFT. For cases where ${N_\mathrm{dof}}\gg N_t$, e.g., the jet, lossless compression and the central difference both contribute to efficiency; for small {$N_\mathrm{dof}$}, e.g., the cavity, the central difference is responsible for most of the advantage.

Comparing Fig.~\ref{fig:scaling}a and b also reveals that for sine-taper SPOD the wall times of the cavity and jet cases are of the same order of magnitude, despite a 26-fold difference in {$N_\mathrm{dof}$}. Clearly the wall time does not scale with grid size. Together with reduced memory usage, we anticipate this feature will be crucial to the analysis of three-dimensional flow data that lack spatial homogeneities.

\section{Lossy compression}\label{sec:appLossy}
\begin{figure}
    \centering
    \includegraphics{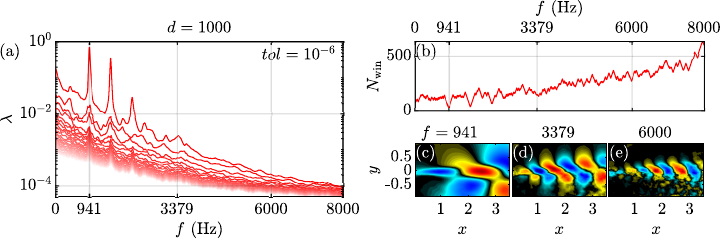}
    \caption{Same as Fig.~\ref{fig:spod_tap2tap} for $tol=10^{-6}$ (red curves), but computed with lossy compression, retaining the first $d=1000$ singular values and vectors of the weighted data matrix}
    \label{fig:cavity_spod_svdCompress}
\end{figure}
As outlined in Sect.~\ref{sec:compression}, if the computational effort required by adaptive SPOD is excessive, SVD truncation can achieve further acceleration at the expense of convergence at high frequencies, which tend to have low energy. However, once $\nwinf$ has been determined adaptively, we recompute the SPOD using the losslessly-compressed data. The final spectrum and modes are thus dictated solely by $\nwinf$. For the SPOD in Fig.~\ref{fig:cavity_spod_svdCompress} computed with $tol=10^{-6}$, we retain the first $d=1000$ singular values and vectors of the weighted data matrix $\vb{X}$. This represents an additional 16-fold compression relative to lossless compression, and drops the wall time from 18 to 2.4 hours. The spectrum, modes, and $\nwinf$ display excellent agreement with those in Fig.~\ref{fig:spod_tap2tap} for the same value of $tol$.

\bibliographystyle{spmpsci}      
\bibliography{refs}   


\end{document}